\documentclass[aps,pra,10pt,twocolumn,superscriptaddress,showpacs,preprintnumbers,amsmath,amssymb, longbibliography]{revtex4-1}

\usepackage{graphicx}
\usepackage{color}
\usepackage{dcolumn}
\usepackage{times}
\usepackage{bm}
\usepackage{verbatim}
\usepackage{mathbbol}
\usepackage{amsmath}
\usepackage{amssymb}
\usepackage{yfonts}
\usepackage{physics}
\usepackage{mathrsfs}
\usepackage{upgreek}
\usepackage[normalem]{ulem} 
\usepackage[caption=false]{subfig}
\usepackage{float}
\definecolor{dark_red}{rgb}{0.75,0,0}
\definecolor{dark_purple}{rgb}{0.75,0,0.75}
\definecolor{dark_blue}{rgb}{0,0,0.75}
\definecolor{dark_green}{rgb}{0,0.60,0}
\usepackage[colorlinks,citecolor=blue,linkcolor=dark_green]{hyperref}

\newcommand{\XCHEM}{\textsc{xchem}}
\newcommand{\NewStock}{\textsc{NewStock}}

\usepackage{array,kantlipsum}
\usepackage[dvipsnames]{xcolor}
\usepackage{tikz}
\usetikzlibrary{fit}
\usetikzlibrary{
  shapes.multipart,
  matrix,
  positioning,
  shapes.callouts,
  shapes.arrows,
  calc}
  
\definecolor{myyellow}{RGB}{245,177,0}
\definecolor{mysalmon}{RGB}{255,145,73}

\usepackage{smartdiagram}
\usesmartdiagramlibrary{additions}
\usetikzlibrary{babel}
\tikzstyle{bag} = [align=center]

\begin{document}

\title{Two-electron interference in two-photon attosecond double ionization of neon}

\newcommand{\UCFPhys}{Department of Physics, University of Central Florida, Orlando, FL 32816, USA.}
\newcommand{\UCFCREOL}{CREOL, University of Central Florida, Orlando, FL 32816, USA.}
\newcommand{\NIST}{National Institute of Standards and Technology, Gaithersburg, MD, USA.}

\newcommand{\UCDavis}{Department of Chemistry, University of California, Davis, California 95616, USA}
\newcommand{\LBNL}{Chemical Sciences Division, Lawrence Berkeley National Laboratory, Berkeley, California 94720, USA}

\author{Siddhartha Chattopadhyay}
\affiliation{\UCFPhys}
\author{Carlos Marante}
\affiliation{\UCFPhys}
\author{Barry I. Schneider}
\affiliation{\NIST}
\author{C. William McCurdy}
\affiliation{\UCDavis}
\affiliation{\LBNL}
\author{Luca Argenti}\email{luca.argenti@ucf.edu}
\affiliation{\UCFPhys}
\affiliation{\UCFCREOL}

\date{\today}

\begin{abstract}
The pump-probe experiments enabled by X-ray free-electron lasers (XFEL) will allow us to directly observe correlated electronic motion with attosecond time resolution by detecting photoelectron pairs in coincidence. In helium, the transition between the non-sequential and sequential regime in two-photon double ionization (TPDI) is well explained by a virtual-sequential model. Much less is known, however, about the TPDI process in more complex atoms. Recently, we extended the virtual-sequential model to arbitrary light pulses [Chattopadhyay {\it et al.,} Phys. Rev. A~{\bf 108}, 013114 (2023)]. This extension employs multi-channel scattering states for the single ionization of both the neutral and the ionized target, which we initally applied to helium. In the present study, we show that our extended virtual-sequential model reproduces the qualitative features of the angularly integrated observables with available experimental results for neon, a considerably more complex target. 
We observe an intriguing feature of inverted two-particle interference in the joint energy distribution of $\mathrm{Ne}$ compared to $\mathrm{He}$. This phenomenon, attributable to the presence of a final doubly ionized state with triplet symmetry coupled to the two photoelectrons, should be observable with current experimental technologies.
\end{abstract}

\maketitle

\section{\label{sec:Introduction}Introduction}
 Our understanding of the attosecond dynamics of photoionization has been revolutionized by the development, in the last two decades, of coherent extreme ultraviolet (XUV) and X-ray light sources~\cite{Corkum-NPhys-07, Krausz-RMP-09, Pazourek-RMP-15, Midorikawa-NPhoto-22}. 
Several effective attosecond pump-probe schemes have been developed to probe time-resolved dynamics in atoms and molecules. Prominent examples are attosecond transient absorption spectroscopy (ATAS)~\cite{Beck-CPL-15, Geneaux-PTRSA-19}, attosecond streaking~\cite{Itatani-PRL-02, Mairesse-PRA-05, Cousin-PRX-17}, and the reconstruction of the attosecond beatings by the interference of two-photon transitions (RABBITT)~\cite{Paul-Science-01, Muller-APB-02}. In some of these schemes, such as streaking and ATAS, the dynamics excited by weak pump XUV pulses is probed by an intense IR field, which alters the dynamics under study. Transformative advances in the generation of strong ultrashort x-ray pulses at Free-Electron-Laser facilities (XFELs), have opened the door to the use of attosecond-pump attosecond-probe techniques to investigate correlated attosecond dynamics in atoms and molecules with unprecedented time resolution~\cite{Duris-NPhoto-20, Saito-Optica-19, Kretschmar-Optica-22, Rolles-APX-22, Borrego-Varillas2022}. 
In addtion, XUV/x-ray pump-probe spectroscopies  can be site-specific and able to provide both high temporal and high spatial resolution of the excitation/ionization of localized core orbitals.
Double ionization is a relevant process inherent to all pump-probe experiments conducted with pairs of ionizing pulses. When photoelectrons are measured in coincidence, double ionization provides unique information on the collective electron dynamics in the sample, due to its high sensitivity to electron correlation~\cite{Weber-Nature-00, Maansson-NPhys-14,bello-PRA-19} and to the entanglement of the photofragments~\cite{Maxwell-NComm-22}.
The sequential regime of two-photon double ionization (TPDI) has been explored by measuring the recoil ion momentum distribution with a free-electron-laser (FEL) facility~\cite{Rudenko-PRL-08, Kurka-JPB-09, Rudenko-JPB-10}. Another study with a high-harmonic generation source was able to distinguish between the non-sequential and sequential regimes by tuning the central photon energy~\cite{Manschwetus-PRA-16}. An experiment performed with the Fermi FEL facility further highlighted the role of autoionizing states in the TPDI of $\mathrm{Ne}$~\cite{Carpeggiani-NPhys-19}. Recently, Orfanos \emph{et. al.} has investigated the TPDI of $\mathrm{Ne}$ with two XUV photons produced at ELI-ALPS facility~\cite{Orfanos-PRA-22}.

Since two-photon double ionization can be interpreted and computed within the framework of time-dependent perturbation theory (TDPT), it is uniquely well-suited to elucidate multi-electron dynamics in multiphoton processes.  Furthermore, in contrast to direct simulations, TDPT allows us to disentangle the contribution from different intermediate ionization pathways. To make headway in this direction, we recently developed a finite-pulse virtual-sequential model (FPVSM) for the two-photon double ionization of polyelectronic atoms~\cite{Chattopadhyay-PRA-23}. As a proof of principle, we used this method to study several angularly integrated observables in the ionization of helium, finding excellent agreement with available time-dependent Schrödinger equation (TDSE) simulations~\cite{Palacios-PRL-09, Feist-PRL-09, Jiang-PRL-15}. To the best of our knowledge, for more complex atoms, the time-resolved dynamics associated with the TPDI process has been investigated for beryllium~\cite{Yip-PRA-15} and magnesium~\cite{Bello-PRA-20} only, within a two-active electron model. A method that accounts for many-body effects in both the neutral and the intermediate ions, such as our FPVSM, offers insight into the TPDI of atoms as complex as neon, both in the sequential (SDI) and in the non-sequential (NSDI) regime, where the two photons are absorbed before the intermediate ion relaxed to a well-defined state. 

It is well known that a single quantum particle can give rise to fringes associated with the interference between alternative quantum paths, as in the famous double-slit experiment. In 1989, Horne~\emph{et. al.}~\cite{Horne-PRL-89} described, for pairs of entangled photons, a similar quantum phenomenon, now known as two-particle interference, where no interference fringes are visible in the distribution of either photons, when considered separately, whereas the coincidence signal does exhibit interference. The spatially separated entangled particles give rise to nonlocal phenomena and have been measured using interferometric techniques in the quantum optics community~\cite{Hochrainer-RMP-22}. This phenomenon, however, is not limited to bosons. Instead, it is expected for any set of identical particles under the appropriate conditions. The interference due to exchange symmetry has no classical counterpart. The two-particle interference between electrons was theoretically predicted by Végh and Macek~\cite{Vegh-PRA-94} in Auger-Photoelectron coincidence spectroscopy and later identified with synchrotron sources in xenon~\cite{Schwarzkopf-JPB-96, Viefhaus-PRL-98}. However, in the absence of time-resolved theoretical descriptions or measurements, some essential features of the two-particle interference are lost. In 2009, Palacios~\emph{et al.} described the signature of two-particle interference in the TPDI of the helium atom, by solving numerically the TDSE for the atom exposed to a sequence of ultrashort ionizing pulses~\cite{Palacios-PRL-09, Palacios-JPB-10}. These studies showed how the ionization probability as a function of the energies $E_1$ and $E_2$ of the two electrons detected in coincidence would give rise to interference fringes parallel to the diagonal $E_2=E_1$. This result is expected for electronic pairs coupled to singlet spin multiplicity, whose spatial part is even under the permutation of the two particles. In a recent work~\cite{Chattopadhyay-PRA-23}, the FPVSM successfully reproduced the characteristic feature of the two-particle interference observed in \emph{ab initio} simulations~\cite{Palacios-PRL-09,Palacios-JPB-10}, which illustrates the predictive potential of the method, despite its modest computational cost. Thanks to advances in the generation and control of attosecond pulses, the possibility of studying two-electron interference with coincident detection is on the horizon. 

In this work, we extend the model to polyelectronic atoms. This extension allowed the study of the joint electron distribution for electronic pairs with arbitrary multiplicity. As predicted by McCurdy~\cite{McCurdy-22}, the presence of a final doubly-charged state with triplet spin multiplicity in two-photon double ionization should give rise to a two-particle interference pattern with inverted troughs and peaks, compared to the ionization to singlet-coupled photoelectron pairs. We confirm this prediction by applying the FPVSM to the TPDI of the neon atom. Furthermore, we have used the FPVSM to reproduce the measurement, by Kurka~\emph{et al.}~\cite{Kurka-JPB-09}, of the ion-resolved single-ultrashort-pulse TPDI of the neon atom in sequential regime, finding a good agreement, confirming the ability of the model to make quantitatively accurate predictions.

The paper is organized as follows. In Sec.~\ref{sec:Theo}, we extended the FPVSM starting from TPDI amplitude derived in Ref.~\cite{Chattopadhyay-PRA-23}.  In Sec.~\ref{sec:CloseCoupling}, we discuss the close-coupling (CC) expansion to compute the bound-continuum transition matrix elements for the neutral and intermediate parent-ion states of neon. In Sec.~\ref{sec:Results}, the FPVSM is used to compute the joint energy  distribution for the ionization of neon by a single XUV pulse, in both the non-sequential and sequential regime. In Sec.~\ref{sec:TPI}, we discuss the two-particle interference  in the presence of the final grand-parent ion with triplet and singlet symmetry. Finally, a two-color pump-probe scheme is proposed to detect the two-particle interference in the TPDI process.   In Sec.~\ref{sec:Conc}, we present our conclusions. Atomic units ($\hbar = 1$, $m_e=1$, $q_e=-1$) and the Gauss system are used throughout unless stated otherwise.

\section{\label{sec:Theo}Theory}
In this section, we illustrate the formulas for the photoelectron joint-energy distribution in the two-photon double-ionization of a polyelectronic atom by means of a sequence of ultrashort pulses, within the approximations entailed by the virtual-sequential model~\cite{Chattopadhyay-PRA-23}.
Since most atomic photoionization studies are conducted on rare-gas targets, we will consider the two-photon double ionization process for an atom $\mathrm{X}$ assumed to be initially in a $^1S$ ground state $g$, giving rise to two electrons with well-defined asymptotic energy, orbital angular momentum, magnetic and spin-projection quantum numbers, 
\[
\mathrm{X}_g + \gamma_1+\gamma_2\longrightarrow \mathrm{X}^{2+}_{A,M_A,\Sigma_A}+e^-_{E_1\ell_1 m_1\sigma_1} + e^-_{E_2\ell_2 m_2\sigma_2}.
\]
where $A$ identifies the electronic state of the doubly-charged ion, with $M_A$ and $\Sigma_A$ being its magnetic and spin-projection quantum numbers.
Within the assumptions of the finite-pulse virtual sequential model~\cite{Chattopadhyay-PRA-23}, the transition amplitude has the following expression in terms of reduced bound-continuum transition matrix elements for the neutral and ionized system and of the external-field parameters,
\begin{eqnarray}
&&\mathcal{A}^{(2)}_{A, E_2 \ell_2 m_2 \sigma_2, E_1 \ell_1 m_1 \sigma_1 \gets g}
=\frac{1-\mathcal{P}_{12}}{2i\sqrt{3}} 
C_{\frac{1}{2}\sigma_2,\frac{1}{2}\sigma_1}^{S_A-\Sigma_A}\sum_{La}\Pi_{LS_A}^{-1}\,
\times\nonumber\\
&&\times\sum_{ij}
\int_{-\infty}^{\infty} d\omega \,
\frac{\tilde{F}_j(E_A+E_1+E_2-E_g-\omega) \tilde{F}_i(\omega)}{E_g+\omega-E_a-E_2+i0^+}
\times\nonumber\\
&&\times\sum_{M M_a\mu\nu}
C_{L_A M_A, \ell_1 m_1}^{L M} 
C_{L_a M_a,\ell_2 m_2}^{1 \mu}
C_{L_a M_a, 1 \nu}^{L M}
\epsilon_{j}^\nu\epsilon_i^\mu\,\times\nonumber\\
&&\times\langle \Psi^{^{2S_a+1}L^{\bar{\pi}_a}(-)}_{A \ell_1 E_1}\|\mathcal{O}_1\|\Phi_{a}\rangle\,
\langle\Psi_{a\ell_2 E_2}^{{^1P^o}(-)}\|\mathcal{O}_1\|g\rangle,
\label{eq:2PDIAmp}
\end{eqnarray}
where $\vec{F}(t)=\sum_i \hat{\epsilon}_i F_i(t)$ is the external field, formed by several independent pulses with fixed polarization, $\tilde{F}_i(\omega)=\frac{1}{\sqrt{2\pi}}\int e^{i\omega t}F_i(t) dt$ are the Fourier Transforms of the individual pulse amplitudes, $\mathcal{P}_{12}$ exchanges all the subsequent indices for photoelectrons 1 and 2, $C_{a\alpha,b\beta}^{c\gamma}$ are Clebsch Gordan coefficients, and $\Pi_{a}=\sqrt{2a+1}$. The state vector $|\Psi_{a\ell E}^{^1P^o(-)}\rangle$ represent a scattering state of the neutral atom, with overall $^1P^o$ symmetry, fulfilling incoming boundary conditions, and whose outgoing component is formed by a parent ion $\mathrm{X}^+$ in the state $|\Phi_a\rangle$ and a photoelectron with orbital angular momentum $\ell$ and energy $E$. Similarly, the state vector $\Psi^{^{2S_a+1}L^{\bar{\pi}_a}(-)}_{A \ell' E'}$ represents a scattering state of the parent ion, with the same multiplicity and opposite parity as $|\Phi_a\rangle$, total angular momentum $L$, fulfilling incoming boundary conditions, and whose outgoing component is formed by a doubly-charged ion $\mathrm{X}^{2+}$ in a state labelled $A$ and a second photoelectron with orbital angular momentum $\ell'$ and energy $E'$. The reduced matrix elements are defined as $\langle\phi_{L'}\|\mathcal{O}_1\|\psi_{L}\rangle = \Pi_{L'}^{-1}\sum_{M,M',\mu} C_{LM,1\mu}^{L'M'} \langle\phi_{L'M'}|\mathcal{O}_{1\mu}|\psi_{LM}\rangle$. In the numerical implementation of these formulas, it is convenient to represent the external field as a combination of Gaussian pulses. Indeed, for Gaussian pulses, the frequency integral can be expressed analytically in terms of the Faddeeva 
function, which can be evaluated numerically at a negligible computational cost. 

Since the virtual-sequential model does not take into account the post-ionization interaction between the two photoelectrons, the amplitudes estimated with \eqref{eq:2PDIAmp} cannot be used to compute the angularly-resolved photoelectron distribution. However, it is well known that the same amplitudes have been shown to produce accurate predictions for the joint-energy photoelectron distribution $dP/dE_1dE_2$~\cite{Chattopadhyay-PRA-23,Horner-PRA-07}, which is obtained by angularly integrating the fully differential distribution and by summing over the spin-projection quantum numbers,
\begin{equation}
\frac{dP_A}{dE_1dE_2}=\sum_{M_A\Sigma_A\sigma_1\sigma_2}\int d\Omega_1 d\Omega_2 \left| \mathcal{A}^{(2)}_{A,E_2\Omega_2\sigma_2,E_1 {l}_1 \Omega_1\gets g}\right|^2,
\end{equation}
where $\Omega_1$ and $\Omega_2$ are the two photoelectron solid emission angles.
The angular integration is equivalent to the summation over the photoelectrons' azimuthal and magnetic quantum numbers. Since the orientation of the doubly-charged ion is normally not measured, we also need to sum over the magnetic and spin quantum number of the ion,
\begin{equation}\label{eq:JED}
\frac{dP_A}{dE_1dE_2}=\sum_{M_A\Sigma_A} \sum_{\{l_im_i\sigma_i\}}\left| \mathcal{A}^{(2)}_{A,E_2{l}_2 m_{2}\sigma_2,E_1 {l}_1 m_{1}\sigma_1\gets g}\right|^2.
\end{equation}
By substituting~\eqref{eq:2PDIAmp} in \eqref{eq:JED}, it is possible to show that the joint energy distribution can be expressed in the following form, which is manifestly symmetric,
\begin{eqnarray}
    \frac{dP_A}{dE_1dE_2}(E_1,E_2)&=&\frac{dP^1_A}{dE_1dE_2}(E_1,E_2)+\frac{dP^1_A}{dE_1dE_2}(E_2,E_1) + \nonumber \\
    &+& \frac{dP^2_A}{dE_1dE_2}(E_1,E_2) + \frac{dP^2_A}{dE_1dE_2}(E_2,E_1).
    \label{eq:TotalTPDIAmp}
\end{eqnarray}
The expressions for the two functions $dP^{1/2}/dE_1dE_2$ can be evaluated using standard techniques of angular-momentum algebra~\cite{Varshalovich-88}. 
For simplicity, we have assumed that all fields are linearly polarized along $\hat{z}$, $\epsilon_i^\mu=\delta_{\mu 0}$ $\forall i$. The result is
\begin{eqnarray}\label{eq:TPDIAmp1}
&&\frac{dP^1_A}{dE_1dE_2}(E_1,E_2)=\frac{1}{4}
\sum_{J=0}^2 (C_{10,10}^{J0})^2 \times\nonumber\\
&\times&\sum_{ab} f_{a}(E_2) f_{b}(E_2)^*
\sum_{L\ell\ell'}
\left\{
\begin{matrix}
\ell&L_a&1\\
L_b&L&1\\
1&1&J
\end{matrix}
\right\} \times\\
&\times&\langle\Psi_{a\ell E_2}^{{^1P^o}(-)}\|\mathcal{O}_1\|g\rangle
\langle\Psi_{b\ell E_2}^{{^1P^o}(-)}\|\mathcal{O}_1\|g\rangle^*
\times\nonumber\\
&\times&
\langle \Psi^{^{2S_a+1}L^{\bar{\pi}_a}(-)}_{A \ell' E_1}\|\mathcal{O}_1\|\Phi_{a}\rangle
\langle \Psi^{^{2S_b+1}{L}^{\bar{\pi}_b}(-)}_{A \ell' E_1}\|\mathcal{O}_1\|\Phi_{b}\rangle^*,\nonumber
\end{eqnarray}
and
\begin{eqnarray}\label{eq:TPDIAmp2}
&&\frac{dP^2_A}{dE_1dE_2}(E_1,E_2)=
\frac{1}{4}
\frac{(-1)^{S_A}}{(2S_A+1)}
\sum_{J=0}^2 (C_{10,10}^{J0})^2 \times\nonumber\\
&\times& \sum_{ab} f_{a}(E_2) f_{b}(E_1)^*\sum_{LL'}
\sum_{\ell_1\ell_2
}
\Pi_{L}\Pi_{L'}\times\nonumber\\
&\times&
\left\{
\begin{matrix}
1 & 1      & J \\
L & \ell_2 & L_a 
\end{matrix}
\right\}
\left\{
\begin{matrix}
\ell_2 & L  & J \\
\ell_1 & L' & L_A 
\end{matrix}
\right\}
\left\{
\begin{matrix}
1  & 1 & J \\
L' & \ell_1 & L_b 
\end{matrix}
\right\} \times\\ 
&\times&\langle \Psi^{^{2S_a+1}L^{\bar{\pi}_a}(-)}_{A \ell_1 E_1}\|\mathcal{O}_1\|\Phi_{a}\rangle\langle\Psi_{a\ell_2 E_2}^{{^1P^o}(-)}\|\mathcal{O}_1\|g\rangle\times\nonumber \\ 
&\times&\langle \Psi^{^{2S_b+1}{L'}^{\bar{\pi}_b}(-)}_{A \ell_2 E_2}\|\mathcal{O}_1\|\Phi_{b}\rangle^*\,
\langle\Psi_{b\ell_1 E_1 }^{{^1P^o}(-)}\|\mathcal{O}_1\|g\rangle^*.\nonumber
\end{eqnarray}
The expressions $\langle \psi \| \mathcal{O}\|\phi\rangle$ indicate reduced dipole matrix elements, the sets of six or nine quantum numbers delimited by curly brackets indicate six-J symbols and nine-J symbols, respectively~\cite{Varshalovich-88}, and the coefficients $f_a(E_n)$ are defined as
\begin{equation}\label{eq:f_a}
f_a(E_n)=\sum_{ij}
\int_{-\infty}^{\infty} d\omega \,
\frac{\tilde{F}_j(E_A+E_1+E_2-E_g-\omega) \tilde{F}_i(\omega)}{E_g+\omega-E_a-E_n+i0^+}.
\end{equation}
As mentioned above, for Gaussian pulses the frequency integrals, $\tilde{F}_i$, can be expressed in closed form in terms of the Faddeyeva function. The detailed formulas are provided in Appendix B of \cite{Chattopadhyay-PRA-23}.
In the simulation of a pump-probe experiment employing a fixed pump pulse $F_1(t)$ and a probe pulse with a controllable delay $\tau$, $F_2(t;\tau)=F_2(t-\tau)$, the Fourier transform of the probe is $\tilde{F}_2(\omega;\tau) = \tilde{F}_2(\omega;0) e^{i\omega \tau}$. Thus, the coefficients $f_a(E_n)$ acquire a parametric dependence on the delay, which is responsible for the interference fringes observed in the joint photoelectron energy distribution.
In particular, when the two Gaussian pulses do not overlap, the non-sequential contribution to the ionization is exponentially suppressed, with the only residual appreciable contribution to the integral coming from the pole in Eq.~\eqref{eq:f_a}. Thus, for $\tau$ sufficiently larger than the duration of the two pulses,
\begin{eqnarray}
f_a(E_1)&\simeq&
\tilde{F}_2(E_A-E_a+E_2) \tilde{F}_1(E_a+E_1-E_g)\times\nonumber\\
&\times& \pi\,\exp\left[i(E_A-E_a+E_2)\tau-i\pi/2\right]\\
f_a(E_2)&\simeq&
\tilde{F}_2(E_A-E_a+E_1)\tilde{F}_1(E_a+E_2-E_g)\times\nonumber\\
&\times& \pi\,\exp\left[i(E_A+E_1-E_a)\tau-i\pi/2\right].
\end{eqnarray}
If we focus on the contribution to the joint photoelectron distribution by a single intermediate ionic channel $a$, the component $dP_A^1/dE_1dE_2$ in \eqref{eq:TPDIAmp1} does not depend on the time delay, since it contains the factor $f_a(E_2)f_a(E_2)^*$. For the same intermediate channel, on the other hand, the component $dP_A^2/dE_1 dE_2$ in \eqref{eq:TPDIAmp2} depends on the pulse delay through a global factor $e^{-i\Delta E\,\tau}$, where $\Delta E\equiv E_2-E_1$. The product of the reduced ionization amplitudes approximately contribute an additional phase factor $e^{i(\tau_{a\ell_2} - \tau_{A\ell_2}) \Delta E +i\Delta_\ell\phi_{a}-i\Delta_\ell\phi_A}$, where $\tau_{a\ell_2}=\partial_E \mathrm{arg}\langle\Psi_{a\ell_2 E}^{{^1P^o}(-)}\|\mathcal{O}_1\|g\rangle$ and $\tau_{A\ell_2}=\partial_E \mathrm{arg}\langle \Psi^{^{2S_a+1}L^{\bar{\pi}_a}(-)}_{A \ell_2 E}\|\mathcal{O}_1\|\Phi_{a}\rangle$ are Wigner photoemission delays in the ionization of the neutral to the $a\ell_2$ channel, and of the ion $a$ to the $A\ell_2$ channel; $\Delta_\ell\phi_a = \mathrm{arg}\langle\Psi_{a\ell_2 E_1}^{{^1P^o}(-)}\|\mathcal{O}_1\|g\rangle-\mathrm{arg}\langle\Psi_{a\ell_1 E_1}^{{^1P^o}(-)}\|\mathcal{O}_1\|g\rangle$ and $\Delta_\ell \phi_A = \mathrm{arg}\langle \Psi^{^{2S_a+1}L^{\bar{\pi}_a}(-)}_{A \ell_2 E_1}\|\mathcal{O}_1\|\Phi_{a}\rangle-\mathrm{arg}\langle \Psi^{^{2S_a+1}L^{\bar{\pi}_a}(-)}_{A \ell_1 E_1}\|\mathcal{O}_1\|\Phi_{a}\rangle$.
This additional phase is expected to be small compared to $2\pi$. Thus, to a good approximation, we have
\begin{equation}\label{eq:InterferenceAnalytical}
\begin{split}
&\frac{dP^2_A}{dE_1 dE_2}(E_1,E_2;\tau) + \frac{dP^2_A}{dE_1 dE_2}(E_2,E_1;\tau) =\\
&=(-1)^{S_A}\,\cos(\Delta E\,\tau)\,g(E_1,E_2),
\end{split}
\end{equation}
where $g(E_1,E_2)$ is a positive quantity. For singlet final states and $\Delta E=0$, this term adds to the background amplitude due to $dP_A^1/dE_1dE_2$, whereas for triplet final states it substracts from it.
Notice that the two-particle interference takes place only when both the energies and the photoemission angles of the electron pair in the first path coincide with those in the second path. This circumstance is automatically taken into account in the present formalism.

\section{\label{sec:CloseCoupling} Electronic-structure calculations.}
This section details the parameters used to compute the multi-configuration bound and continuum states of the neutral neon target and of the intermediate singly-ionized system, Ne$^+$, and it compares the results for the energy and the photoionization cross sections with experimental values from the literature.

\begin{figure}[t]
\includegraphics[width=\linewidth]{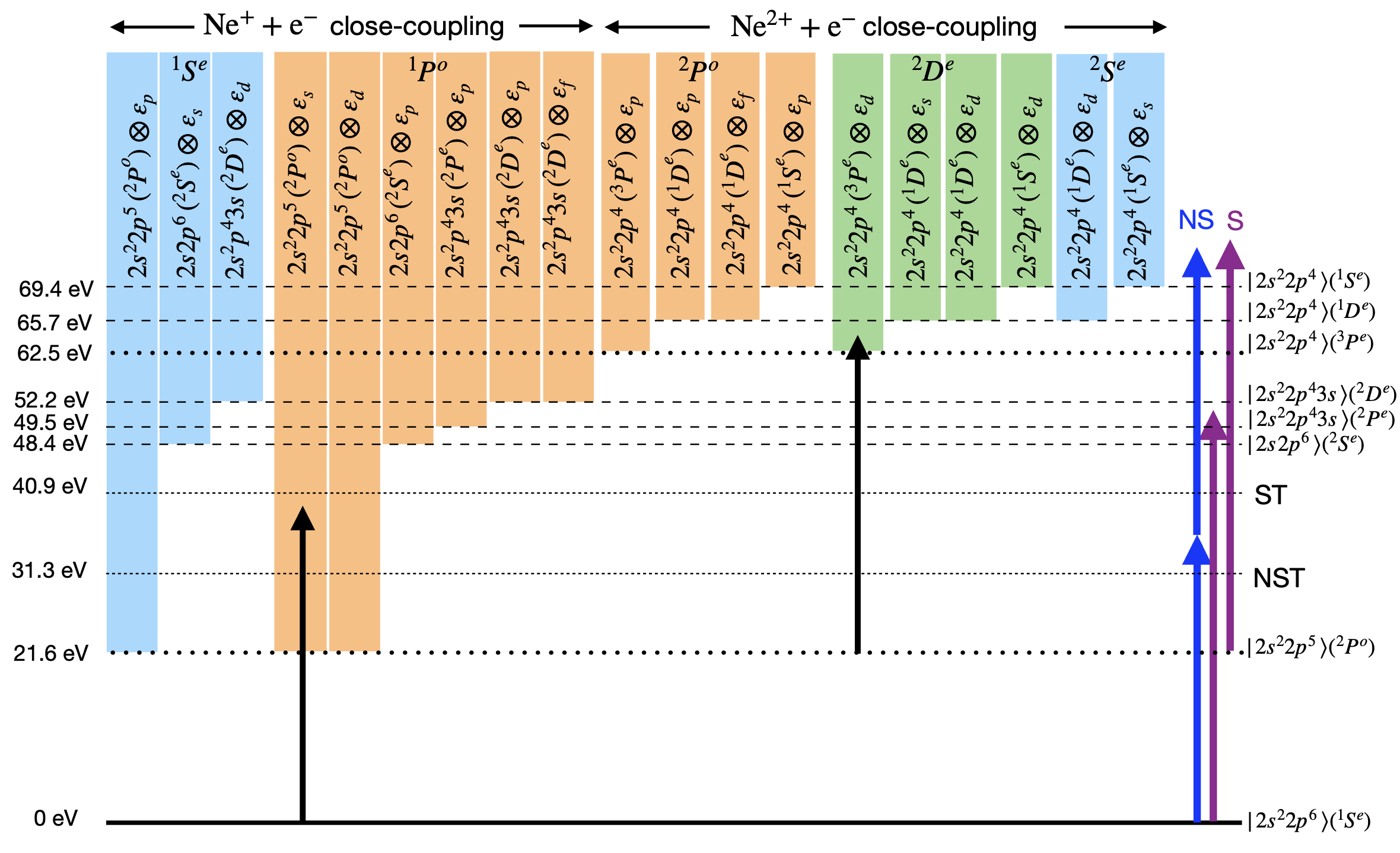}
\caption{\label{Fig:TPDINeEnScheme} Energetics and close-coupling scheme for two-photon double ionization of $\mathrm{Ne}$. The total symmetries, $S$, $P$, and $D$ are indicated by light blue (online), orange (online), and green (online) of neutral and the ionized atom. Black arrows indicate direct one-photon ionization from the ground state of the respective system. The non-sequential regime (dark blue, online) dominates when two XUV photons with energy 31.3~eV~$<\omega<$~40.9~eV doubly ionize the $\mathrm{Ne}$ atom.  At energies well above the sequential threshold, $\omega>$~40.9~eV, the sequential mechanism (purple, online) dominates.} 
\end{figure}

Both the bound and the continuum states of the system are obtained using the close-coupling (CC) with pseudochannel method implemented in the \NewStock{}  atomic photoionization code~\cite{Carette-PRA-13}. The energetics and associated CC expansion scheme for the TPDI  of $\mathrm{Ne}$ is shown in Fig.~\ref{Fig:TPDINeEnScheme}. In the case of a single pulse, or a sequence of identical pulses, the non-sequential mechanism dominates when the XUV photon energy is between 31.3 eV and 40.9 eV. Beyond the sequential threshold of 40.9 eV, the TPDI process is dominated by the sequential absorption of two XUV photons.

In contrast to the TPDI of $\mathrm{He}$, the $\mathrm{Ne}^{2+}$ ion produced in the TPDI of $\mathrm{Ne}$ exhibits states with different multiplicity: the triplet ground state, $1s^22s^22p^4\,({^{3}P^e})$, and two excited singlet states, $1s^22s^22p^4\,({^{1}D^e})$, $1s^22s^22p^4\,({^{1}S^e})$, which are 3.3~eV and 6.9~eV above the ground level, respectively. This circumstance gives rise to a richer spectrum and it allows us to recognize the permutational symmetry of the spatial photoelectron wave packet in the two-particle interference of the ion-resolved joint-energy distribution of the two photoelectrons, as discussed in Sec.~\ref{sec:TPI}.

To build the CC space for the single ionization of the neutral atom, we need to specify both the set of localized configurations that are needed to reproduce electronic correlation at short range, the set of correlated parent ions used in the CC expansion, and the single-particle space for the ejected photoelectron. At the energies considered in this work, the two $1s$ core electrons do not participate to the dynamics, nor does their correlation play any appreciable role in the ionization. For the present work, therefore, we assume that the $1s$ shell is doubly occupied and frozen. For simplicity, in the following we will omit the $1s^2$ part of the atomic configurations. The configuration-interaction (CI) space for the $\mathrm{Ne}^+$ parent ion is generated by all parity-preserving single and double excitations from the reference configurations $2s^22p^5$ (odd states) and $2s2p^6$ (even states) to $n=3$ orbitals. The states and the orbitals are optimized with the state-averaged Multi-Configuration Hartree-Fock method (MCHF) that is part of the ATSP2K package~\cite{Fischer-CPC-07}. The parent-ion energies are listed in Table~\ref{tab:ParentIonEn}.

\begin{table}[hbtp!]
\caption{ Comparison of $\mathrm{Ne}^{+}$ energies (in eV) with NIST data.}
\label{tab:ParentIonEn}
\begin{ruledtabular}
\begin{tabular}{lccc}
        Configuration & \NewStock{}  &  NIST~\cite{NIST_ASD} & $\Delta{E}$  \\
  \hline  
  \hline           
 $2s^2 2p^5$ \quad ($^2P^o$)            & 0.000   &  0.000  & 0.000 \\ 
 $2s 2p^6$  \hspace{0.4cm} ($^2S^e$)    & 26.918  &  26.910 & 0.008 \\ 
 $2s^2 2p^4 3s$ ($^2P^e$)               & 28.619  &  27.859 & 0.760 \\ 
 $2s^2 2p^4 3s$ ($^2D^e$)               & 31.553  &  30.549 & 1.004 \\ 
\end{tabular}
\end{ruledtabular}
\end{table}
To correctly represent the ionization continuum, the CC expansion must include all the optically accessible channels that are open at the energy of interest. In this work, we consider photons with energy up to $49$~eV. This energy is above the excitation threshold of the first four $\mathrm{Ne}^+$ doublet states, which have dominant configuration $2s^2 2p^5(^2P^o)$,  $2s 2p^6(^2S^e)$,  $2s^2 2p^4 3s (^2P^e)$, and  $2s^2 2p^4 3s(^2D^e)$. For this reason, the close-coupling space comprises all the channels generated by these states. The partial wave channels (PWC) are obtained by augmenting the set of ionic states with single-electron states with orbital angular momentum $\ell\leq 3$, giving rise to neutral states with either $^1S^e$ or $^1P^o$ symmetry. The radial part of the photoelectron orbitals are expanded in a B-spline basis of order 7, within a quantization box with a radius of 300~a.u., and with an asymptotic node separation of 0.4~a.u. Past calculations have shown that, with this choice, the resonance parameters computed with \NewStock{} are in excellent agreement with the experimental data~\cite{Puskar-PRA-23}. The PWCs are complemented by the set of localized configurations obtained by adding one electron to the full set of configurations used to generate the parent ions. This additional set of states, known as pseudo channels, ensures that correlation is treated consistently at short and long range. The partial-wave channels are schematically shown in Fig.~\ref{Fig:TPDINeEnScheme}. 
\begin{figure}[b!]
\includegraphics[width=\columnwidth]{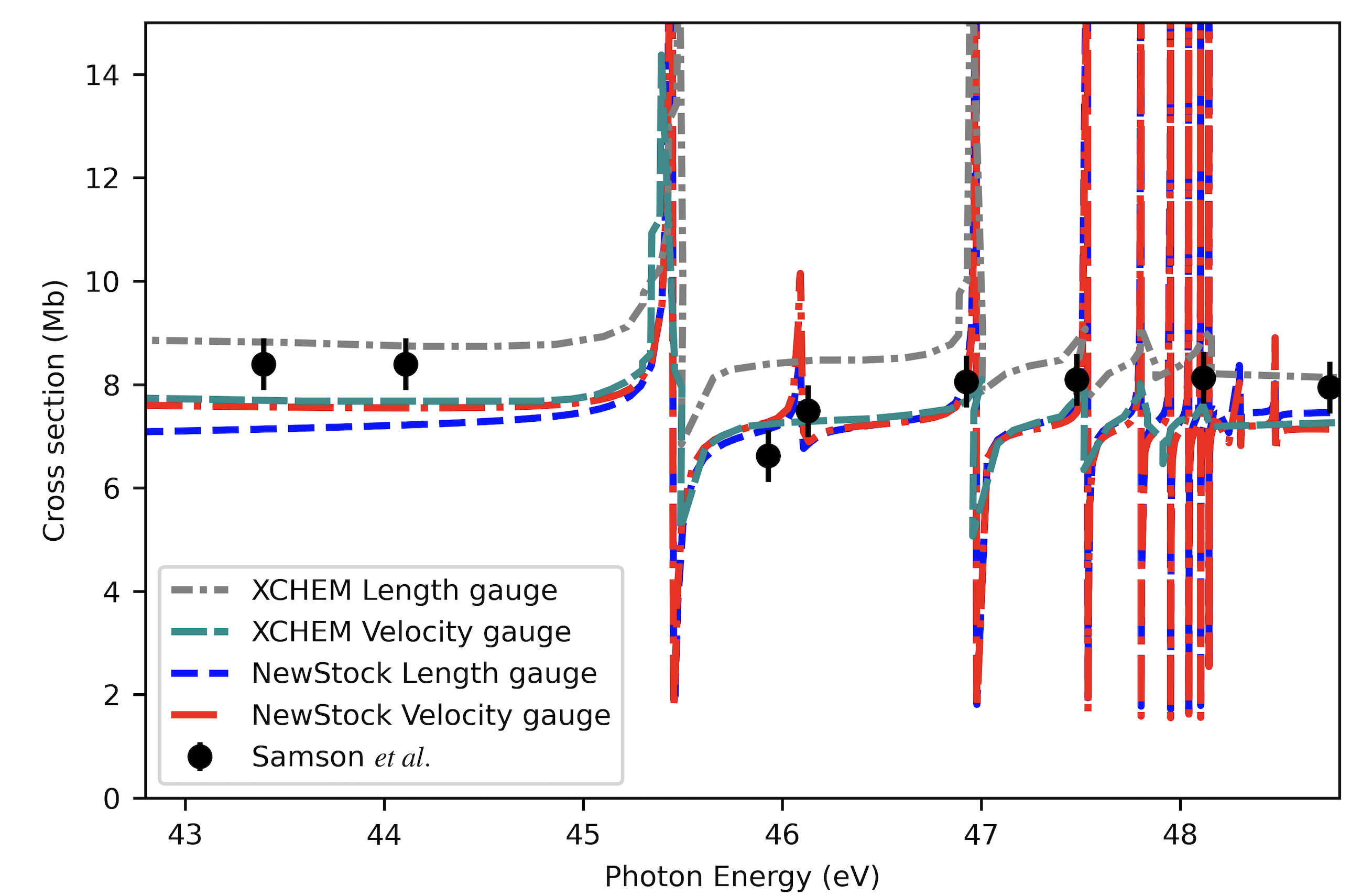}
\caption{Total photoionization cross-section from the $\mathrm{Ne}$ ground state computed with \NewStock{}  in length (blue, online) and velocity gauge (red, online) compared with the length (grey, online) and in velocity gauge (darkcyan, online) results obtained from \XCHEM{} ~\cite{Marante-PRA-17} and experimental results~\cite{Samson-02} (black, online). Results are between the $2s^2 2p^5$ ($^2P^o$) and $2s 2p^6$ ($^2S^e$) thresholds. See text for details.}
\label{Fig:NeICrossSection}
\end{figure}
To ascertain the validity of the bound-continuum dipolar couplings, in Fig.~\ref{Fig:NeICrossSection} we compare the total one-photon single-ionization cross-section of $\mathrm{Ne}$ computed in this work with existing experimental results~\cite{Samson-02} as well as with theoretical predictions obtained with \textsc{xchem}~\cite{Marante-PRA-17}, in the energy region between 43~eV and 49~eV, which comprises the second single-ionization threshold. The few experimental data available in this region agree very well with our background cross section. The spectrum features several resonant peaks. The position of those that correspond to the $2s 2p^6 np$ autoionizing  series are in excellent agreement with the \XCHEM{} calculations. The calculations in the present work, however, shows also a resonance around 46.2~eV, due to an intruder state ($2s^2 2p^4 3s3p$) associated with the $^2P^e$ closed channel. The resonance is not reproduced by the spectrum computed with \XCHEM{} in~\cite{Marante-PRA-17} since in that work the $^2P^e$  channel was not included.

The procedure followed to simulate the second ionization step, $\mathrm{Ne}^{+}_a + \gamma \rightarrow \mathrm{Ne}^{2+}_A + e^-$, coincides with the one that was used for the ionization of the neutral. The $\mathrm{Ne}^{2+}$ CI space comprises all single and double excitations from $2s^22p^4$ and $2s2p^33s$ to $n=3$ orbitals. The grand-parent ionic states and the orbitals are determined with a state-average MCHF calculation on the three states with dominant configuration $2s^22p^4$, whose energies are listed in Table~\ref{tab:GrandParentIonEn}.
\begin{table}[hbtp!]
\caption{ Comparison of $\mathrm{Ne}^{2+}$ energies (in eV) with NIST data.}
\label{tab:GrandParentIonEn}
\begin{ruledtabular}
\begin{tabular}{lccc}
        Configuration & \NewStock{}    &  NIST~\cite{NIST_ASD} & $\Delta{E}$  \\
  \hline  
  \hline           
 $2s^2 2p^4$ ($^3P^e$)    & 0.000 &  0.000  & 0.000 \\
 
 $2s^2 2p^4$ ($^1D^e$)    & 3.315 &  3.204  & 0.111 \\ 
 
 $2s^2 2p^4$ ($^1S^e$)    & 6.919 &  6.912  & 0.007 \\ 
 
\end{tabular}
\end{ruledtabular}
\end{table}
The PWC are obtained by augmenting these three ionic states with the same single-electron orbitals used for the neutral, to give rise to states with $^2S^e$ or $^2D^e$ symmetry, as sketched in Fig.~\ref{Fig:TPDINeEnScheme}.
\begin{figure}[htb]
\includegraphics[width=\columnwidth]{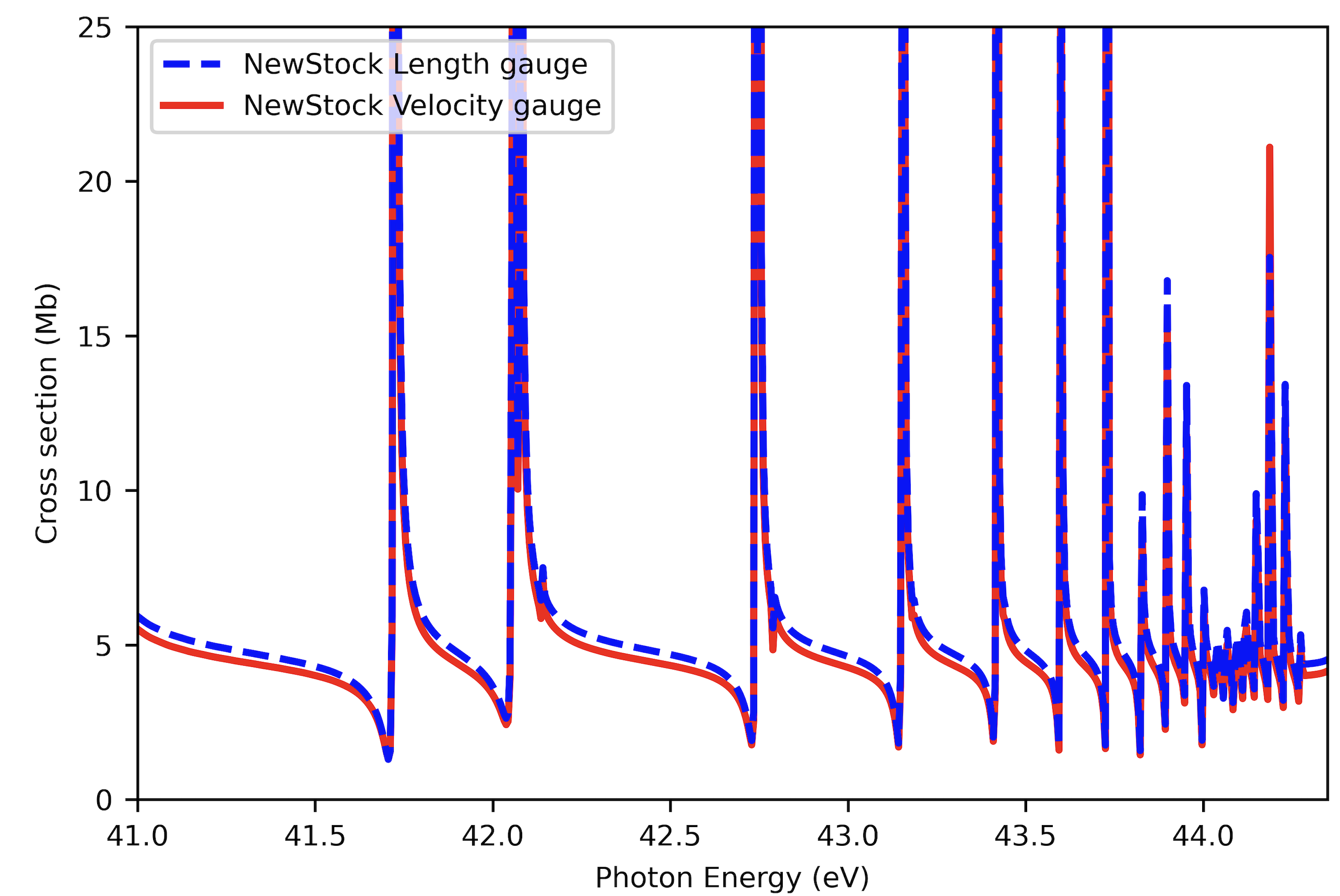}
\caption{Total photoionization cross-section from the $\mathrm{Ne}^{+}$ ground state ($^2P^o$) computed with \NewStock{}  in length ( blue, online) and velocity gauge (red, online).  Results are between the $2s^2 2p^4$ ($^3P^e$) and $2s^2 2p^4$ ($^1D^e$) thresholds. See text for details.}
\label{Fig:NeIICrossSection}
\end{figure}
Figure~\ref{Fig:NeIICrossSection} shows the present predictions for the one-photon single-ionization cross-section from the $^2P^o$ ground state of $\mathrm{Ne}^{+}$. The spectrum is dominated by the $2s^2 2p^4(^1D^e)n{d}$ and $2s^2 2p^4(^1S^e)ns$ autoionizing  series. The results obtained in length and velocity gauges are in even better agreement than in the ionization of the neutral. This result is expected, since, when the charge increases, the correlation energy converges to a finite value~\cite{Bartell1965} and hence the ratio between the correlation energy and the electronic excitation energy decreases, thus making the configuration expansion converge faster. In the present work, all the TPDI observables are computed in length gauge.

\section{\label{sec:Results}One-pulse two-photon double ionization}
\begin{figure}[b!]
\includegraphics[width=\columnwidth]{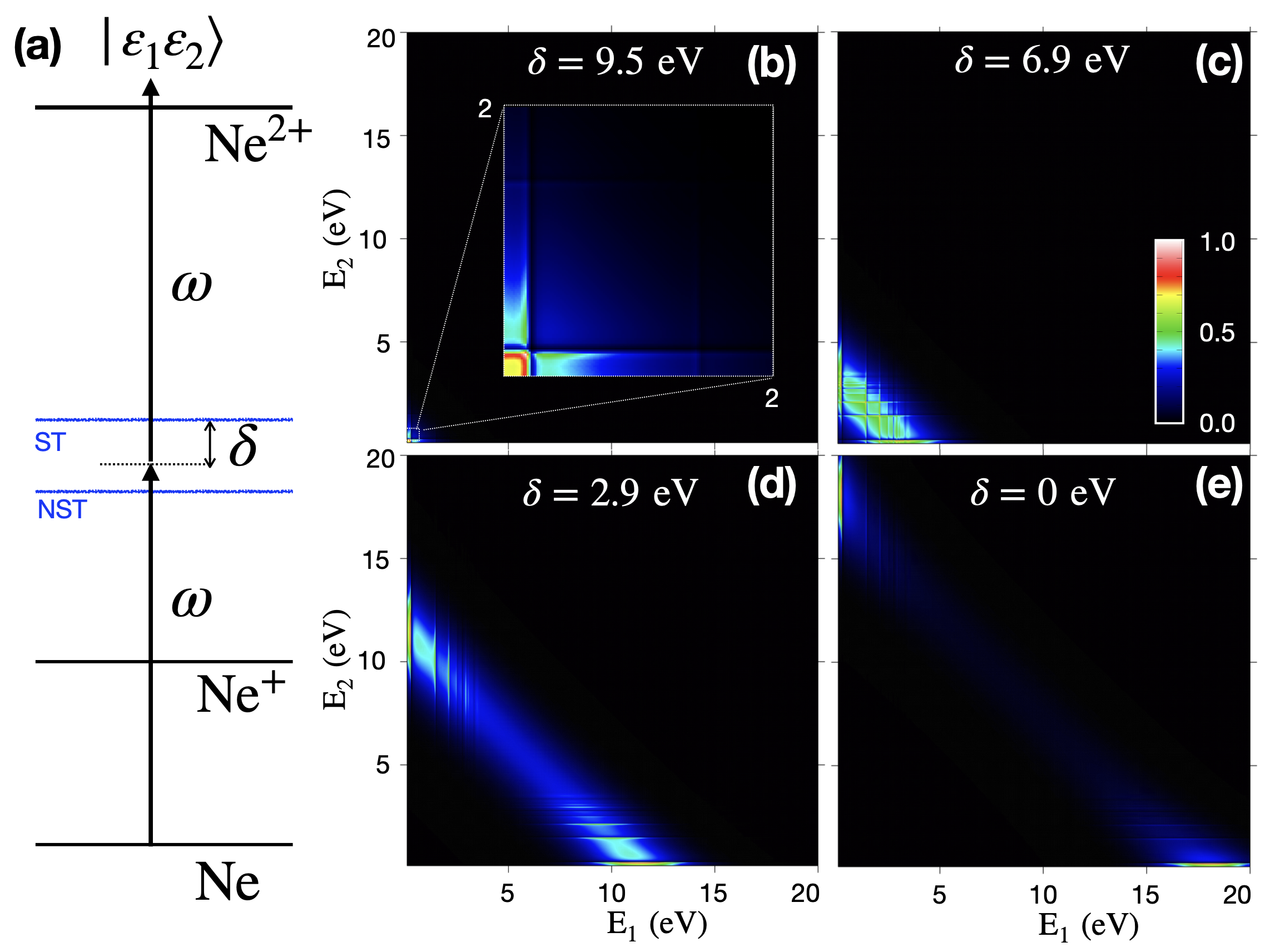}%
\caption{(a) TPDI scheme for $\mathrm{Ne}$ in the non-sequential regime. $\delta$ is the energy difference from the sequential threshold (ST). Joint energy distribution of two photoelectrons in the non-sequential regime with a fixed pulse duration of 500~as and intensity of $4\times10^{10}$~W/cm$^2$ for different central photon energies:  (b) 31.5 eV, (c) 34 eV, (d) 38 eV, and (e) 40.9 eV. See text for details.}
\label{Fig:JED_NS_Ne}
\end{figure}

In this section, we present our results from the FPVSM starting from the opening of the non-sequential threshold at 31.3 eV and up to the sequential threshold of 40.9 eV to illustrate the features of the electron-correlation-dominated non-sequential regime. In this regime, the intermediate states play a significant role, as in the case of $\mathrm{He}$. We use 500~as Gaussian pulses for the simulation with a peak intensity of $4\times10^{10}$~W/cm$^2$.  Fig.~\ref{Fig:JED_NS_Ne}~(a) shows the TPDI scheme in the non-sequential regime with $\delta$ as the energy difference from the sequential threshold. As shown in Fig.~\ref{Fig:JED_NS_Ne}~(b), at the opening of the non-sequential threshold, the signal emerges at $\delta$~=~9.5 eV. At the central photon energy $\omega$~=~34 eV, the joint energy distribution exhibits a broad range of energy sharing, as shown in Fig.~\ref{Fig:JED_NS_Ne}~(c). This unique feature, which is similar to what was already observed in $\mathrm{He}$~\cite{Chattopadhyay-PRA-23,Feist-PRA-08}, is a hallmark of TPDI processes below the sequential threshold.  Additional sharp features appear in the joint photoelectron energy distribution. As commented in~\cite{Chattopadhyay-PRA-23}, these are artifacts of the FPVSM due to the fact that, in our model, the intermediate ionic state is available instantaneously, even if the lifetimes of the autoionizing resonances may be longer than the pulse duration. As we approach the sequential threshold of 40.9 eV, the two pronounced peaks at 19.3~eV  appear as shown in  Fig.~\ref{Fig:JED_NS_Ne}~(e). Close to the sequential threshold of 40.9 eV and above,  this characteristic two-peak structure suggests that an independent-particle model is sufficient to explain the joint photoelectron energy distribution.   

\begin{figure}[t!]
\includegraphics[width=\columnwidth]{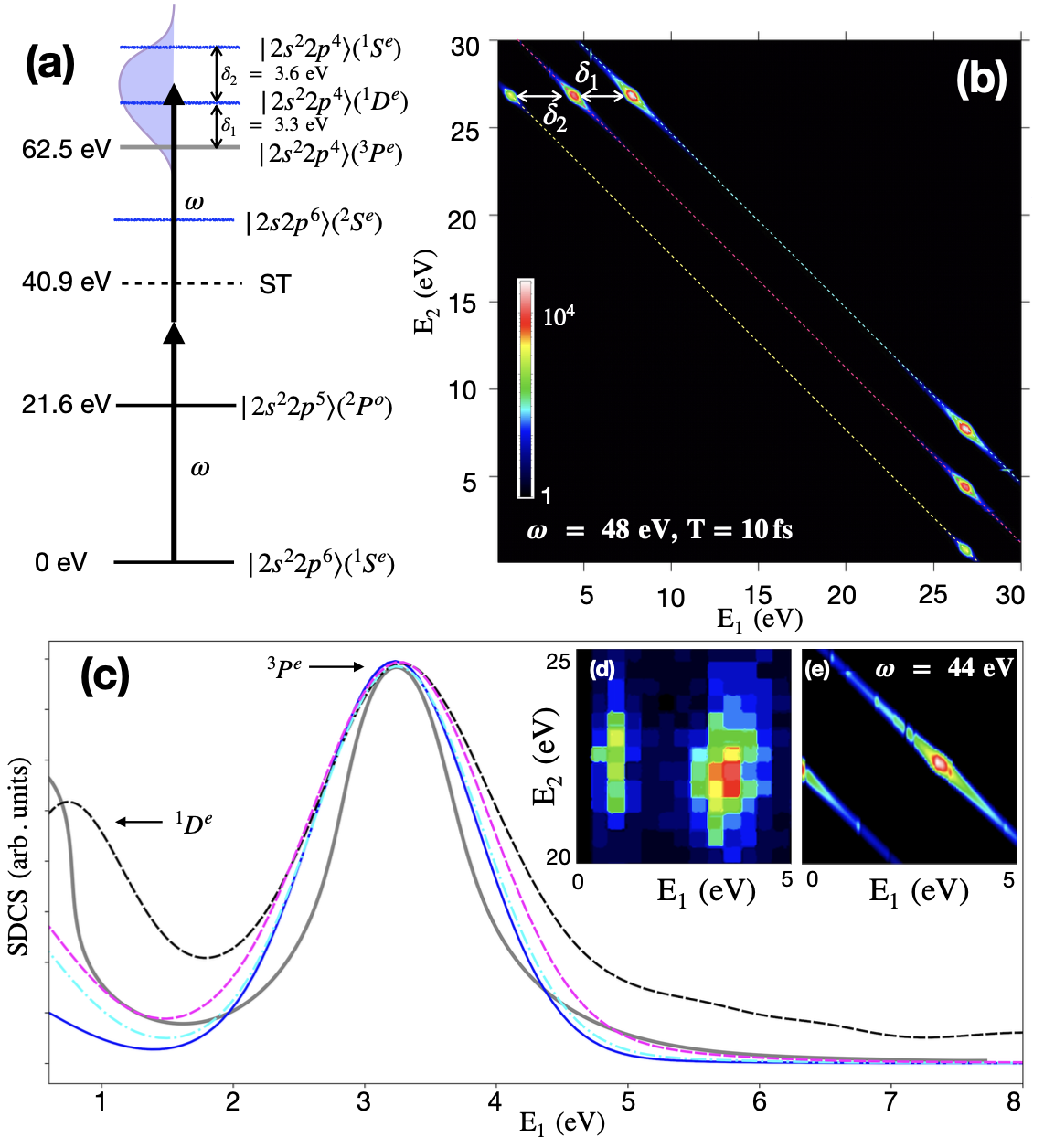}%
\caption{(a) TPDI scheme for $\mathrm{Ne}$ in the sequential regime. $\delta_1$ is the energy difference between the $2s^2 2p^4(^3P^e)$ ground-state of the $\mathrm{Ne}^{2+}$ and the excited state $2s^2 2p^4(^1D^e)$, and $\delta_2$ is the separation between $2s^2 2p^4(^1D^e)$ and $2s^2 2p^4(^1S^e)$ excited states. (b) Joint energy distribution of two photoelectrons in the sequential regime with a fixed pulse duration of 10 fs and intensity of $4\times10^{10}$~W/cm$^2$ with a central photon energy of 48~eV. (c) Comparison of the singly-differential cross-section with Ref.~\cite{Kurka-JPB-09} for different pulse duration. The grey curve is the experimental result, and the black curve is the energy-integrated singly-differential cross-section obtained from Fig.~1(a) of Ref.~\cite{Kurka-JPB-09}. The blue curve is the corresponding simulation with a pulse duration of 25~fs. 25~fs (blue solid), 10~fs (cyan dash dotted), 5~fs ( magenta dashed). (d) Digitized section of experimental result from Fig.~1~(a) of Ref.~\cite{Kurka-JPB-09} (e) Simulation with the same set of laser parameters as provided in Ref.~\cite{Kurka-JPB-09}. See text for details.}
\label{Fig:JED_Seq_Ne}
\end{figure}

In the case of XUV-photon energies above the sequential threshold, several open intermediate channels dominate the sequential regime, giving rise to rich dynamics following TPDI.  Contrary to $\mathrm{He}^{2+}$, the multiplicity of the grand-parent ionic ground state of $\mathrm{Ne}^{2+}$ is different. It has triplet symmetry, and the two ionic excited states with the same configuration $2s^2 2p^4$ belong to singlet symmetry as shown in Table~\ref{tab:GrandParentIonEn}. From a theoretical perspective, the signal corresponding to the multiple grand-parent ionic states in the joint energy distribution in the presence of a single pulse, therefore, is a unique characteristic of any model.  Figure~\ref{Fig:JED_Seq_Ne}~(a) shows the sequential TPDI scheme to probe the final three grand-parent ionic states of $\mathrm{Ne}^{2+}$ with different multiplicities with a suitable XUV or X-ray pulse. To investigate the presence of multiple grand-parent ionic states, we use a laser pulse with a central energy of 48~eV and a duration of 10~fs,  and the result is shown in Fig.~\ref{Fig:JED_Seq_Ne}~(b). This choice of laser parameters is dictated by the CC expansion of the ionized atom, and from a measurement perspective, all three grand-parent ionic states can be probed with the current XFEL facilities. The FPVSM produces a qualitative picture with all three grand-parent ionic states as shown in Fig.~\ref{Fig:JED_Seq_Ne}~(b). The  three pairs of peaks correspond to three different final grand-parent ionic states, which are prominent in the joint photoelectron energy distribution. The leading signal appears at $E_{\rm{tot}}$= 35.5~eV and belongs to the $^3P^e$ ground grand-parent ionic state, with a peaked value two-orders of magnitude larger than the signal corresponding to the $^1S^e$ state. The separation of the peaks $\delta_1$ and $\delta_2$ is consistent with the energy difference between the $^3P^e$ and $^1D^e$ state and between the $^1D^e$ and $^1S^e$ state. To visualize the signals for all three grand-parent ionic states, we use a logarithmic scale. The dotted lines highlight the different states along the total energy axis. 

In a 2009 experiment, Kurka \emph{et al.} measured the ion-resolved ($^3P^e$ and $^1D^e$) photoelectron signal in the two-photon double ionization of Neon using a free-electron-laser pulse with central energy of 44~eV, overall duration of 25~fs, coherence duration of 5$\pm1$~fs, and intensity of $5\times10^{13}$~W/cm$^2$~\cite{Kurka-JPB-09}. Figure ~\ref{Fig:JED_Seq_Ne}~(c) compares the experimental results with our theoretical predictions for the singly-differential cross section, computed for different durations of the XUV pulse. We digitized the experimental results in two different ways: i) by digitizing the singly-differential signal reported in the original paper (thick grey line); ii) by digitizing the 2D color plot of the signal in the joint energy distribution, and subsequently numerically integrating the result along the $E_2$ coordinate. In the latter case, the area of the peak at 3.2~eV is as much as three times larger than the peak near threshold. Our simulation with a pulse duration of 25~fs gives a much smaller proportion for the peak at lowest energy, because the width of the spectrum is evidently not sufficient to appreciably reach above the threshold of the excited ion. On the other hand, the 25~fs pulse is not Fourier limited. Indeed, the coherence duration of $5$~fs indicates that the spectral width of the pulse is larger. When a pulse with a duration of 5~fs is used, the first peak increases appreciably in size. Further reduction of the pulse duration does not improve the agreement and it distorts the profile of the main peak at 3.2~eV. In absence of a detailed knowledge of the spectral profile of the pulse, which may itself feature peaks, it is difficult to draw further conclusions on the quantitative agreement between theory and experiment. Given the circumstances, therefore, a branching ratio agreement within a factor of 2 from the experiment is to be expected.
Inset panels~\ref{Fig:JED_Seq_Ne}~(d,e) compare the measured (d) and theoretical (e) signal in the joint photoelectron energy distribution.  Given the absence of clear error bars in the experiment, the qualitative agreement of our numerical results with the experiment highlights the semi-quantitative predictive power of FPVSM in the ion-resolved sequential TPDI regime.

\section{\label{sec:TPI}Pump-probe TPDI}
While there are numerous studies of quantum interference in single-ionization attosecond spectroscopy~\cite{Lindner-PRL-05}, the analogous two-particle interference in double photoionization, has not been studied.  The current lack of such experiments is possibly a consequence of the difficulty of producing  attosecond pulses with sufficient flux to measure XUV-pump XUV-probe coincidence observables. The effort to improve these measurements, however, is justified both by the direct information on correlated electronic motion and by its ability to detect effects that critically depend on the permutational symmetry of the two photoelectrons. In the case of $\mathrm{He}$, TDSE simulations show clear interference fringes in the joint-electron distribution, due to two-particle interference~\cite{Palacios-PRL-09}, which the FPVSM reproduces well~\cite{Chattopadhyay-PRA-23}. 
\begin{figure}[b!]
\includegraphics[width=\columnwidth]{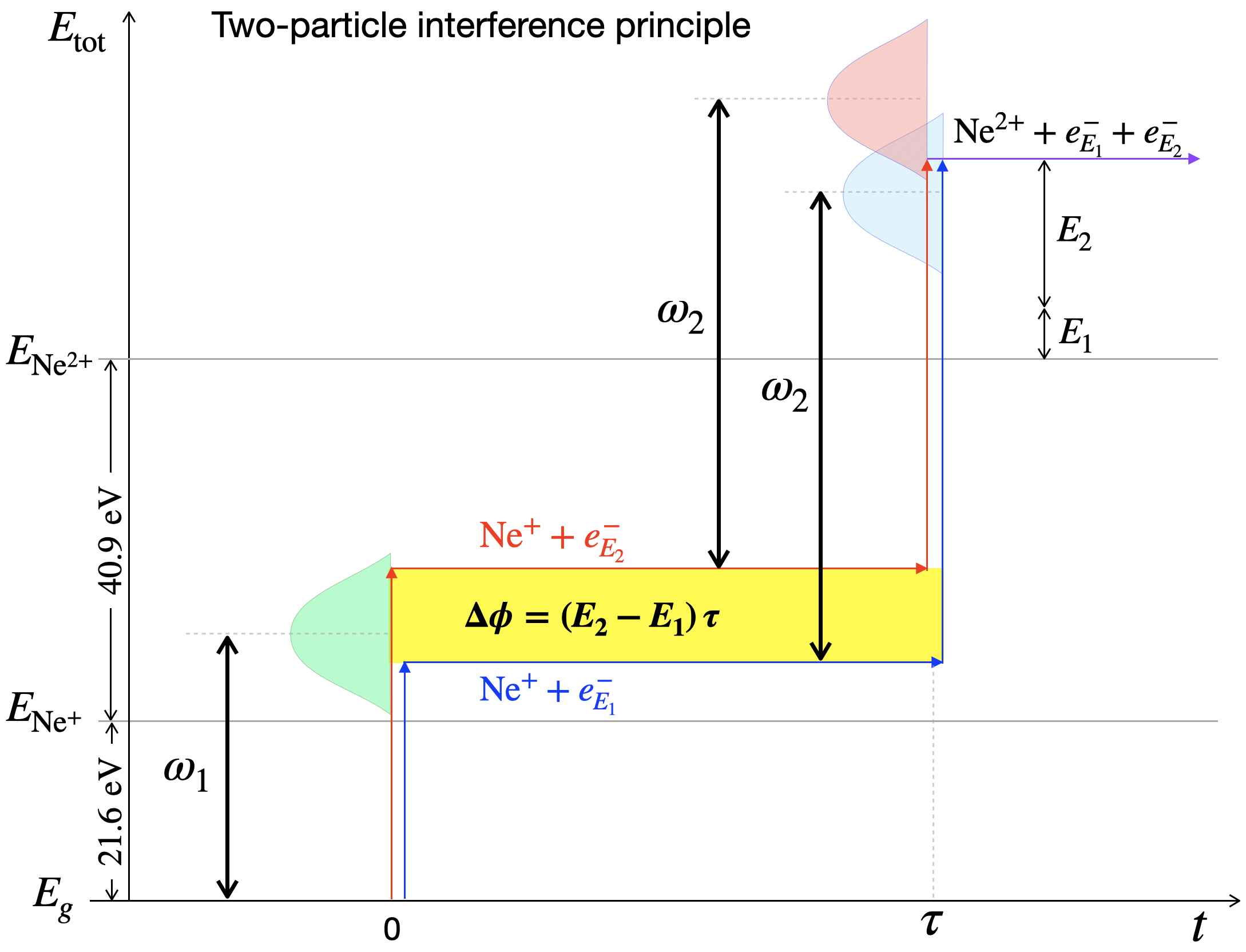}%
\caption{\label{Fig:PumpProbeTPDI} (a) Two-electron quantum interferometric scheme: At time $t$~=~0, a first XUV photon ionizes the neutral $\mathrm{Ne}$ atom. Due to the finite with of the pulse spectrum, both photoelectrons with energy $E_1$ (path 1, blue online) and $E_2$ (path 2, red online) are generated. After a delay $\tau$, a second XUV photon ionizes $\mathrm{Ne}^{+}$. Once again, the finite spectral with generates a distribution of energies for the second electron illustrated by the shaded blue (path 1) and red (path 2) bell shapes, which overlap. If the energy of the second photon is such that, in the overlap region, the total photoelectron energy equals the sum $E_1$ and $E_2$ of the two electrons ejected by the first pulse, the two paths lead to the same final state, but differing in the order with which the electron with energy $E_1$ and the electron with energy $E_2$ are emitted. These two paths interfere constructively or destructively depending on whether their relative phase $\Delta\phi=(E_2-E_1)\tau$ is an even or odd multiple of $\pi$.}
\end{figure}

In Fig.~\ref{Fig:PumpProbeTPDI}, we illustrate the scheme to study the two-particle interference in the TPDI process. Here, we consider a pump-probe scheme with two XUV pulses with a delay $\tau$,
\begin{equation}
    \vec{\mathcal{E}}(t) = \vec{\mathcal{E}}_{\textsc{XUV}_1}(t) + \vec{\mathcal{E}}_{\textsc{XUV}_2}(t-\tau),
\end{equation}
where $\vec{\mathcal{E}}_{\textsc{XUV}_{1/2}}(t)$ indicate the transverse electric fields of the two pulses. Due to the finite spectral width of the pulses, the final triplet state can be reached through multiple pathways. For the schematic representation, we have shown only two paths leading to the final triplet ground state of the grand-parent ion. The different paths (red and blue arrows, online) produce photoelectrons with energies $E_1$/$E_2$ depending on the ionization from the lower or upper-frequency edge of the pump/probe pulse.  In the interval between the two pulses, the energy of the system along the two paths differ by $\Delta E = E_2-E_1$, and hence the two paths acquire a phase difference $\Delta \phi=(E_2-E_1)\tau$.
Because of the same final state for the two paths, the associated amplitudes interfere constructively or destructively if $\Delta \phi$ is an even or odd integer multiple of $\pi$, respectively [see Eq.\eqref{eq:InterferenceAnalytical}].

\begin{figure}[b!]
\includegraphics[width=\columnwidth]{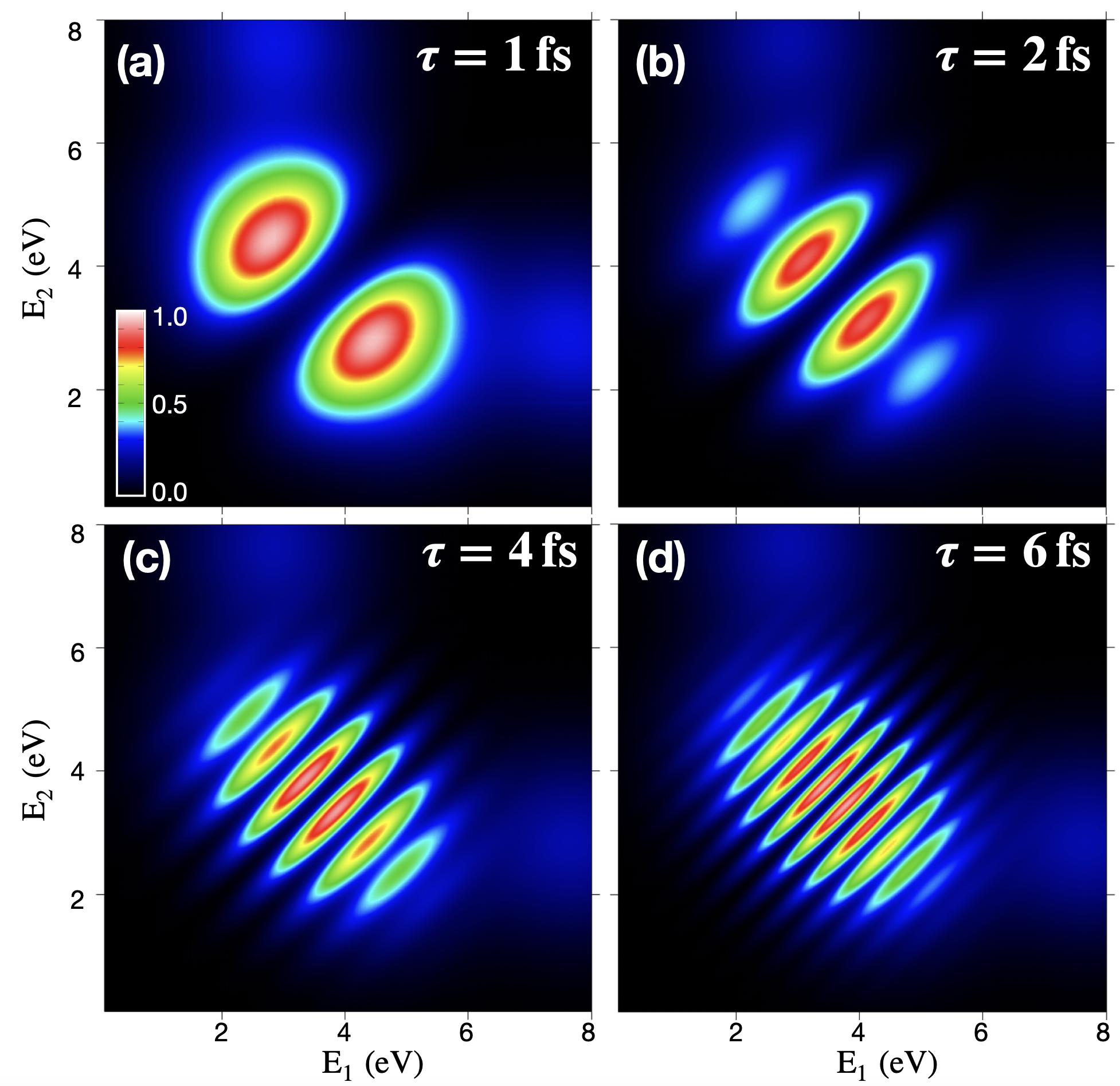}%
\caption{\label{Fig:PumpProbeTPDI_Ne_3Pe} Joint photoelectron energy distribution as a function of the pump-probe delay, in the presence of only $2s^2 2p^4(^3P^e)$ grand-parent ionic state in CC-expansion of the intermediate $2s^2 2p^5(^2P^o)$ ion. Results are shown for different time-delays: (a) $\tau = 1$~fs, (b) $\tau = 2$~fs, (c) $\tau = 4$~fs, and (d) $\tau = 6$~fs. See text for details.}
\end{figure}
To demonstrate the features of the two-particle interference, we use two pulses with central energies $\omega_1=24$~eV and $\omega_2=48$~eV. We chose two specific CC schemes to study the two-particle interference to highlight the presence of the final grandparent ionic state with different multiplicity. In our first scheme, the CC expansion for the target $\mathrm{Ne}^{2+}$-ion includes the $2s^2 2p^4 (^3P^e)$ final grand-parent ionic state. With only the final grand-parent ionic state with triplet multiplicity coupled to the $d$-wave in the CC expansion from the intermediate ion of $2s^2p^5 (^2P^o)$, the results of the joint energy distribution for different time delays are shown in Fig.~\ref{Fig:PumpProbeTPDI_Ne_3Pe}~(a-d). The photoelectron joint energy distribution shows inverted interference fringes compared to what was found in the case of $\mathrm{He}$~\cite{Palacios-PRL-09, Chattopadhyay-PRA-23}. The constructive interference appears for $E_2~\neq~E_1$, and the destructive interference occurs for $E_2=E_1$.  This compelling feature of the inversion of the fringes is explained by Eq.~\eqref{eq:InterferenceAnalytical} 

\begin{figure}[t!]
\includegraphics[width=\columnwidth]{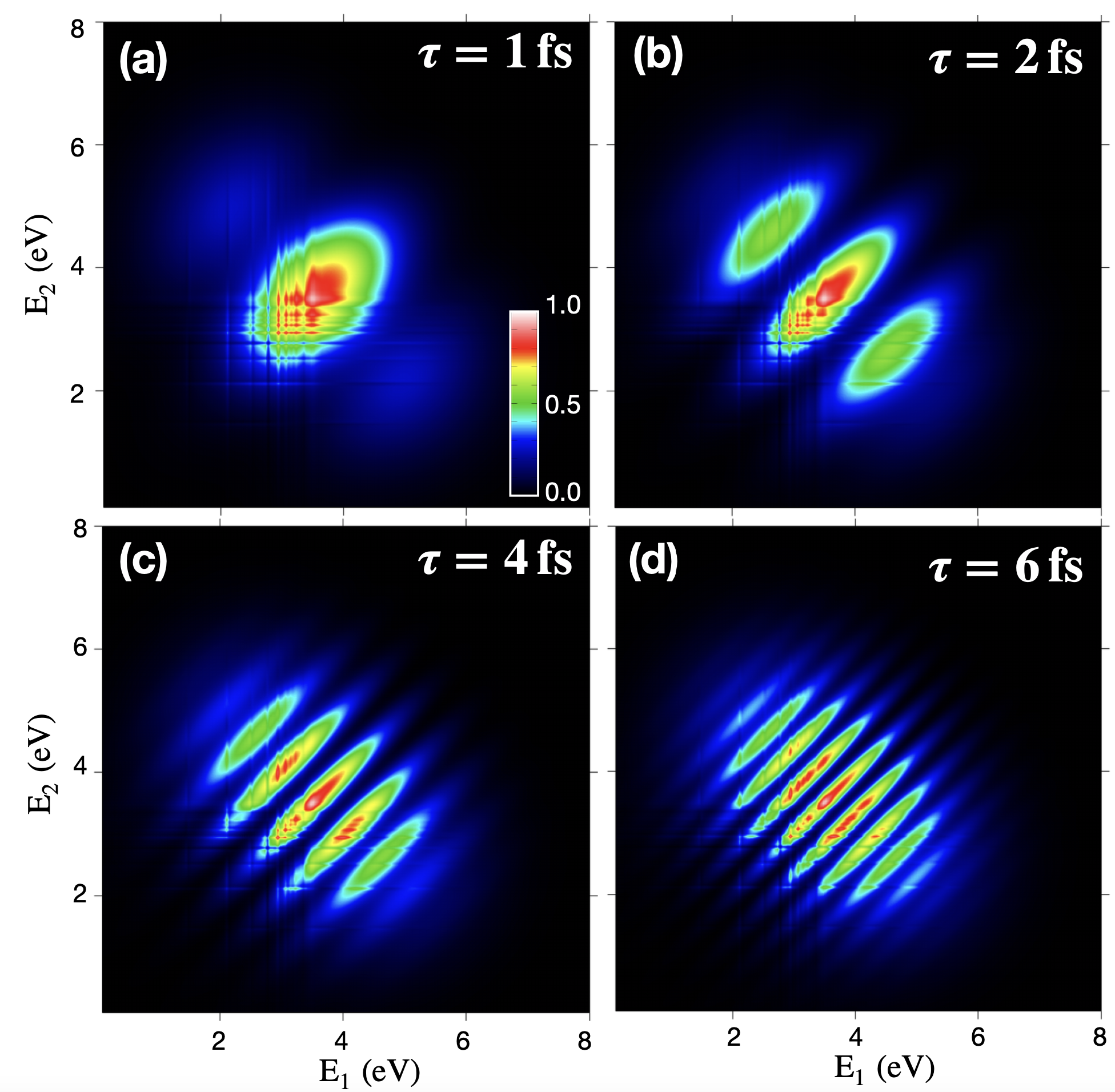}%
\caption{\label{Fig:PumpProbeTPDI_Ne_1Se} Joint photoelectron energy distribution as a function of the pump-probe delay in the presence of only $2s^2 2p^4(^1S^e)$ grand-parent ionic state in the CC expansion of the intermediate $2s^2 2p^5(^2P^o)$ ion. Results are shown for different time-delays: (a) $\tau = 1$~fs, (b) $\tau = 2$~fs, (c) $\tau = 4$~fs, and (d) $\tau = 6$~fs. See text for details.}
\end{figure}
In the second case, the CC expansion for the target $\mathrm{Ne}^{2+}$-ion includes only the $2s^2 2p^4 (^1S^e)$  grand-parent ionic state coupled to the $s$-wave. The corresponding results of the joint energy distribution for different time-delay are shown in Fig.~\ref{Fig:PumpProbeTPDI_Ne_1Se} (a-d). Remarkably, the interference fringes are inverted for different time delays compared to the previous case. The constructive interference appears for $E_2=E_1$, and destructive interference occurs for $E_2~\neq~E_1$ similar to the case of $\mathrm{He}$. Similarly, this feature of the inversion of the fringes can be explained using Eq.~\eqref{eq:TotalTPDIAmp}. However, the spin of the grand-parent ion, $S_A$,  appears only in the third term (Eq.~\eqref{eq:TPDIAmp2}). To illustrate, let us consider the ionization of $\mathrm{Ne}$ to the $2s^22p^4$ ($^1$S$^e$) ionic state by a sequence of two consecutive pulses, which gives rise to a two-particle interference with a maximum along the diagonal $E_1~=~E_2$.  In this case, the prefactor, $\frac{(-1)^{(1-S_A)}}{(2S_A+1)}$ of Eq.~\eqref{eq:TPDIAmp2} become positive and thus the maximum of the interference appears along $E_1=E_2$. Thus, FPVSM shows the two-particle interference pattern in the TPDI process can be used as a tool to detect the spin state of the final grand-parent ion.  

\begin{figure*}[!hbtp]
\includegraphics[width=0.95\textwidth]{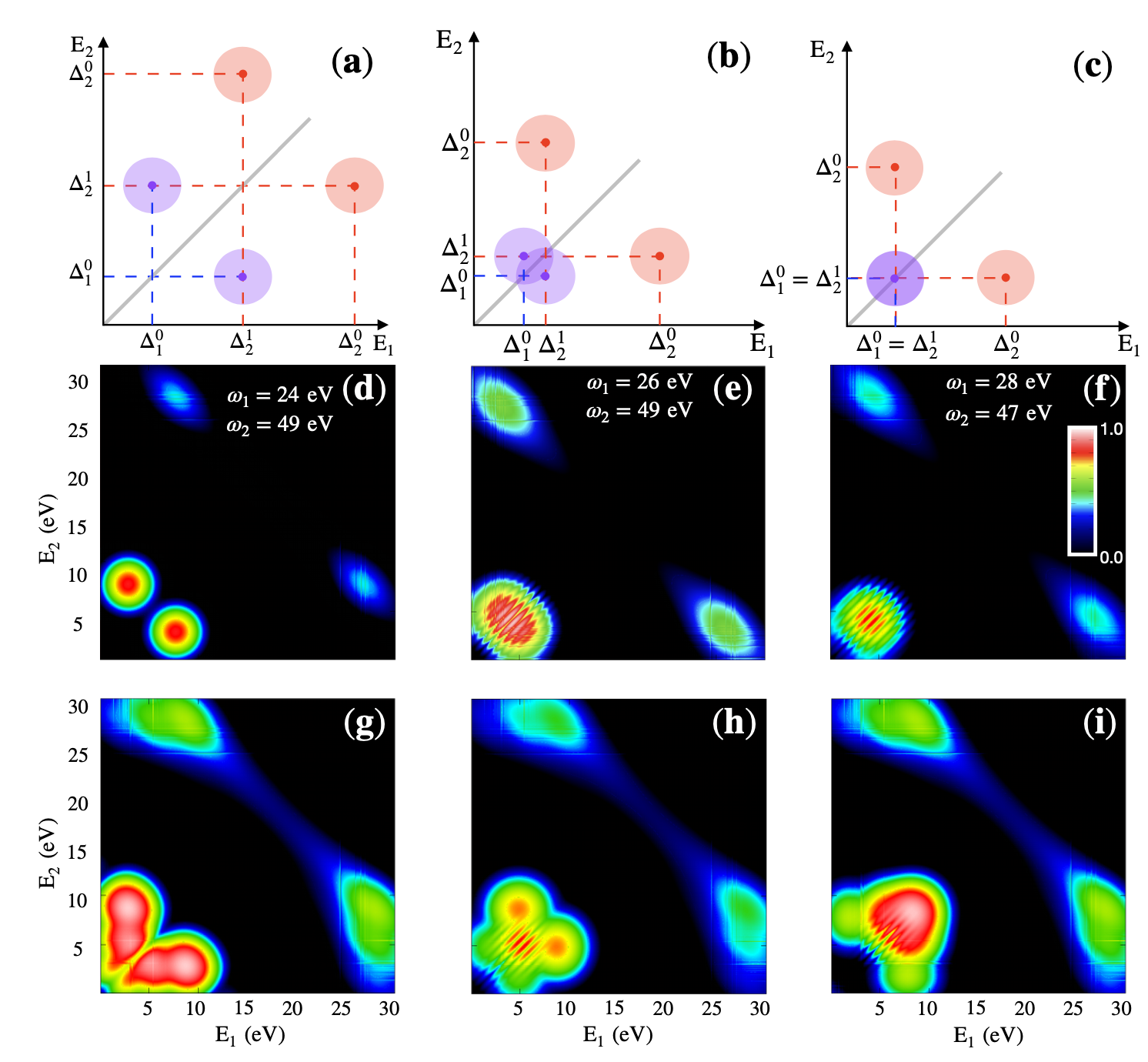}%
\caption{\label{Fig:TPI_TPDI_Ne} Two-particle interference in the joint photoelectron energy distribution in TPDI pump-probe spectroscopy employing three choices of photon energies, ~(24 eV, 49 eV), ~(26 eV, 49 eV), and ~(28 eV, 47eV). (a-c) Scheme to observe two-particle interference in TPDI. Here $\Delta_{1}^{0}=\omega_1-\mathrm{IP}_{\mathrm{Ne}}$, $\Delta_{2}^{1}=\omega_2-\mathrm{IP}_{\mathrm{Ne}^{+}}$, and $\Delta_{2}^{0}=\omega_2-\mathrm{IP}_{\mathrm{Ne}}$; $\omega_1$, $\omega_2$ are the central photon energies, and $\mathrm{IP}_{\mathrm{Ne}}$, and $\mathrm{IP}_\mathrm{Ne}^{+}$ are the ionization potential of the neutral and the intermediate ion respectively. (d-f) Joint photoelectron energy distribution with three different pairs of central frequencies in the presence of only $^3P^e$ final grandparent ionic state. (g-i) Same joint photoelectron energy distribution in the presence of all three grandparent ionic states, $^3P^e$, $^1D^e$, and $^1S^e$.      (a,~d,~g) Joint photoelectron energy distribution with no interference: $\Delta_{1}^{0} < \Delta_{2}^{1} < \Delta_{2}^{0}$.  (b,~e,~h), (c,~f,~i) Joint photoelectron energy distribution with two-particle interference regime: $\Delta_{1}^{0} \le \Delta_{2}^{0} < \Delta_{2}^{0}$. Interference fringes appears as $\Delta_{1}^{0}$ approaches $\Delta_{2}^{1}$.  All the simulations are performed with a time delay, $\tau=4$~fs, and intensity of $4\times10^{10}$~W/cm$^2$. See text for details.}
\end{figure*}
Let us now focus on the two-color pump-probe scheme to observe the two-particle interference that includes the  $2s^2 2p^4(^3P^e)$ grand-parent ionic state in the CC expansion of the intermediate $2s^2 2p^5(^2P^o)$-ion. Based on this CC expansion, we propose a scheme employing three pairs of central frequencies as shown in Fig.~\ref{Fig:TPI_TPDI_Ne}~(a-c) to illustrate the nature of the quantum indistinguishably in determining the two-electron interference in the TPDI process. So far, in the case of $\mathrm{He}$ as well as $\mathrm{Ne}$, we considered specific laser parameters of duration a few hundreds of attoseconds to a few femtoseconds to simulate the two-particle interference and associated dynamics with variable pump-probe delay. A suitable range of laser parameters is required to get more insight into quantum interference, as producing XUV pulses with attosecond duration is still challenging. In the pump-probe TPDI process, the first ionization occurs by absorption of an XUV photon to produce a photoelectron with energy $\Delta_{1}^{0}=\omega_1-\mathrm{IP}_{\mathrm{Ne}}$. After a time delay $\tau$, the second ionization event occurs by absorption of another XUV photon, which can create a photoelectron with energy $\Delta_{2}^{1}=\omega_2-\mathrm{IP}_{\mathrm{Ne}^{+}}$. However, the atom and the intermediate ions can absorb both photons from the first or second pulse and vice versa. Thus, in Fig.~\ref{Fig:TPI_TPDI_Ne}~(a,d), the signals corresponding to 2.5~eV appear when the first low energy photon is absorbed by the neutral atom and the other peak at 8.5~eV appears when the intermediate ion absorbs the high energy photon. The other pair of peaks at 27.5~eV appears when the neutral atom absorbs the high-energy photon. There is no interference in this specific case, because the signals corresponding to the different transition paths are well separated. Indeed, the total amplitude, as expressed in Eq.~\eqref{eq:2PDIAmp} is a linear combination of the product of the dipole amplitudes between the different steps of ionization. As the energy difference between the pump and probe central frequency is larger than the difference between the ionization potentials of the neutral atom and the intermediate ion, the two-photon amplitudes don't overlap and result in a well-separated pair of peaks in the joint photoelectron energy distribution. Thanks to the finite spectral width of the XUV pulses, as $\Delta_{1}^{0}$ approaches $\Delta_{2}^{1}$, the interference fringes appear due to the overlap of the two-photon amplitudes in the joint photoelectron energy distribution as shown in Fig.~\ref{Fig:TPI_TPDI_Ne}~(c,f). As $\Delta_{1}^{0}=\Delta_{2}^{1}$, the two-particle interference is visible due to the pairs of interfering pathways created by the sequence of pulses. Indeed, this feature of two-particle interference in the joint photoelectron energy distribution is a valuable starting point for an experimental measurement. 

\section{\label{sec:Conc} Conclusions}

The present work demonstrates the predictive power of the  FPVSM~\cite{Chattopadhyay-PRA-23} in the pump-probe two-photon double-ionization of neon, as a representative of polyelectronic atoms. The method reproduces well the main features of the joint-energy distribution in the TPDI of $\mathrm{Ne}$ in the non-sequential regime. In the sequential regime, the model correctly predicts the individual contributions to the photoelectron signal from different intermediate ionic states observed in Kurka \emph{et al.}'s 2009 experiment~\cite{Kurka-JPB-09}, thus validating the utility of this model for complex targets. The accuracy of this method for polyelectronic atoms opens the way to its application to the TPDI of molecular targets using only molecular single-ionization methods, such as the one recently implemented in \textsc{astra} code~\cite{Randazzo-PRR-23}.

The approximations in FPVSM allows the modeling and proposal of a practical two-electron interference experiment using neon as the target. Our simulations show the emergence of a clear two-particle interference in the joint photoelectron distribution when the associated ion is left in either a singlet or a triplet state. The interference pattern in the triplet channel has inverted trough and peaks, compared to the singlet channel, which reveales the underlying permutational symmetry of the spatial part of the photoelectron-pair wave function. Remarkably, the three channels with different spin couplings are not predicted to obscure the two-electron interference, even though they have opposite interference patterns, which indicates that this phenomenon should be observable with existing technology.

\begin{acknowledgments}
LA, SC, and CM acknowledges support by the U.S. DOE CAREER Contract No. DE-SC0020311. LA acknowledges support by NSF grant No. PHY-1912507. 
Work by C.W.M. at LBNL was performed under the auspices of the US DOE under Contract DE-AC02-05CH11231 and was supported by the U.S. DOE Office of Basic Energy Sciences, Chemical Sciences, Geosciences, and Biosciences Division. We express our gratitude to Jeppe Olsen for many useful discussions.\end{acknowledgments}


%

\end{document}